\newread\testifexists
\def\GetIfExists #1 {\immediate\openin\testifexists=#1
	\ifeof\testifexists\immediate\closein\testifexists\else
        \immediate\closein\testifexists\input #1\fi}
\def\epsffile#1{Figure: #1} 	

\immediate\openin\testifexists=e
	\ifeof\testifexists\immediate\closein\testifexists\else
        \immediate\closein\testifexists\input e\fipsf 
  
\magnification= \magstep1	
\tolerance=1600 
\parskip=5pt 
\baselineskip= 5 true mm \mathsurround=1pt
\font\smallrm=cmr8

\font\medrm=cmr9

\font\bigbf=cmbx12
 	\def\Bbb#1{\setbox0=\hbox{$\tt #1$}  \copy0\kern-\wd0\kern .1em\copy0} 
	\immediate\openin\testifexists=a
	\ifeof\testifexists\immediate\closein\testifexists\else
        \immediate\closein\testifexists\input a\fimssym.def 
\def\secbreak{\vskip12pt plus .7in \penalty-200\vskip 0pt plus -.5in} 
\def\hugeskip{\vskip12mm plus 3mm}
\def\Narrower{\par\narrower\noindent}	
\def\Endnarrower{\par\leftskip=0pt \rightskip=0pt} 
\def\br{\hfil\break}	\def\ra{\rightarrow}		
\def\a{\alpha}          \def\b{\beta}     \def\G{\Gamma}
\def\d{\delta}          \def\D{\Delta}  
\def\h{\eta}              \def\l{\lambda}          
\def\m{\mu}             \def\f{\phi}                
\def\n{\nu}                 
         \def\s{\sigma}

 \def\LL{{\cal L}} \def\OO{{\cal O}}

\def\cl{\centerline}    
\def\ni{\noindent}             \def\dd{{\rm d}}        
\def\tl{\tilde}                 \def\bra{\langle}       \def\ket{\rangle}
 
\def\fn#1{\ifcase\noteno\def\fnchr{*}\or\def\fnchr{\dagger}\or\def
	\fnchr{\ddagger}\or\def\fnchr{\rm\S}\or\def\fnchr{\|}\or\def
	\fnchr{\rm\P}\fi\footnote{$^{\fnchr}$} 
	{\scrunch#1\toe}\ifnum\noteno>5\global\advance\noteno by-6\fi
	\global\advance\noteno by 1}
 	\def\scrunch{\baselineskip=11 pt \medrm}
 	\def\toe{\vphantom{$p_\big($}}
	\newcount\noteno

\def\ffract#1#2{{\textstyle{#1\over#2}}}
\def\fract#1#2{\raise .35 em\hbox{$\scriptstyle#1$}\kern-.25em/
	\kern-.2em\lower .22 em \hbox{$\scriptstyle#2$}}

\def\half{\ffract12} 

\def\part#1#2{{\partial#1\over\partial#2}} 
 \def\ref#1{${\vphantom{)}}^#1$}

\def\bbf#1{\setbox0=\hbox{$#1$} \kern-.025em\copy0\kern-\wd0
        \kern.05em\copy0\kern-\wd0 \kern-.025em\raise.0433em\box0}              

\def\ref#1{${\,}^{\hbox{\smallrm #1}}$}

\def\Gbar{\raise.13em\hbox{--}\kern-.35em G}
\def\lap{\setbox0=\hbox{$<$}\,\raise .25em\copy0\kern-\wd0\lower.25em\hbox{$\sim$}\,}
\def\glt{\setbox0=\hbox{$>$}\,\raise .25em\copy0\kern-\wd0\lower.25em\hbox{$<$}\,}
\def\gap{\setbox0=\hbox{$>$}\,\raise .25em\copy0\kern-\wd0\lower.25em\hbox{$\sim$}\,}
\def\nc{\medskip\noindent}
\def\iz{\quad = \quad}
\def\dys{\displaystyle}
{\ }\vglue 1truecm
\rightline{THU-98/31}
\rightline{hep-th/9808113}
\hugeskip
\cl{\bigbf COUNTING PLANAR DIAGRAMS}\medskip
\cl{\bigbf WITH VARIOUS RESTRICTIONS}

\hugeskip

\cl{Gerard 't Hooft }
\bigskip
\cl{Institute for Theoretical Physics}
\cl{University of Utrecht, Princetonplein 5}
\cl{3584 CC Utrecht, the Netherlands}
\smallskip\cl{e-mail: \tt g.thooft@phys.uu.nl}
\hugeskip
\ni{\bf Abstract}\Narrower
Explicit expressions are considered for the generating functions concerning
the number of planar diagrams with given numbers of 3- and 4-point vertices.
It is observed that planar renormalization theory requires diagrams with
restrictions, in the sense that one wishes to omit `tadpole' inserions and
`seagull' insertions; at a later stage also self-energy insertions are to 
be removed, and finally also the dressed 3-point inserions and the dressed 
4-point insertions. Diagrams with such restrictions can all be counted exactly.
This results in various critical lines in the $\lambda$-$g$ plane, where
$\lambda$ and $g$ are effective zero-dimensional coupling constants. These lines
can be localized exactly.
\Endnarrower
\hugeskip

{\ni\bf 1.  Introduction}

\nc In Quantum Chromodynamics (QCD), with $N$ color degrees of freedom and
gauge coupling constant $g$, the limit $N\ra\infty$ and $g\ra 0 $, such
that $ N\,g^2=\tl g^2$ is kept fixed, leads to a $\tl g$ expansion, for
which all feynman diagrams are planar\ref1.  The $1/N$ expansion, at higher
orders in $N$, requires diagrams on planes with non-trivial topology.

In this paper, we only consider the $\tl g$ expansion at infinite $N$.  If $\tl g$ is
sufficiently large, this expansion diverges.  Since the theory is
asymptotically free, $\tl g$ is small in the ultraviolet region, so that,
there, small diagrams dominate.  An important question is, how the infrared
region is affected by the fact that the $\tl g$ expansion has a finite
region of convergence.  Naively, one might expect an infrared Landau ghost,
but it is more likely that a delicate rearrangement mechanism will take
place that will strongly affect the physical spectrum of states.  One might
conjecture that this rearrangement mechanism may cause quark confinement in
the $N\ra\infty$ limit.ref2

To this end, we study the critical effects for large $\tl g$'s in a
zero-dimensional model.  The `field theory' is described by the action
$$S(M)={\rm Tr}\left(-\ffract12 M^2+\ffract13 g M^3+\ffract14 \l M^4\right),
\eqno(1.1)$$ where $M$ is an $N\times N$ dimensional matrix, in the linit
$$N\ra\infty\ ,\qquad g,\,\l\ra 0 ,\qquad N\,g^2=\tl g^2\ \hbox{and}\ 
N\l=\tl\l\ \hbox{fixed.}\eqno(1.2)$$ 

Henceforth, the tilde ($\tl{\,}$) will
be omitted.  The main body of the paper consists of the calculation of the
generating functions for the number of planar diagrams with various types
of restrictions on them.  The motivation for the restrictions stems from a
special renormalization program that is particularly suitable for planar
QCD.  This program is deferred to Appendix A, so as not to interrupt the
most important theme of the paper.

We study many different cases, all related one to another by exact equations.  It is
important to have a consistent notation.  Unfortunately, existing notations
were too haphazard, so we had to invent our own.  This is explained in
Section~2.

Much of this paper is based on pioneering work by Koplik, Neveu and
Nussinov\ref3 -- which in turn makes use of earlier work by Tutte\ref4 --
and on the work by Br\'ezin, Itzykson, Parisi and Zuber\ref5.  The
latter apply matrix theory to do the integration with the action~(1.1).
Here, we choose, instead, to work directly from the functional equations,
as these will be easier to handle in the QCD case, and they are also more
transparent in diagrammatic approaches.  The relations are read off 
directly from the diagrams.

A delicate problem then is the choice of boundary conditions for these
equations.  They can be derived by carefully considering the holomorphic
structure that the generating functions are required to have.  Once this is
understood, a fundamental solution is obtained for the generating function
describing {\it the numbers of all planar diagrams for all multiparticle
connected Green functions, with given numbers $V_3$ of three-point
vertices, $V_4$  four-point vertices, and $E$ external lines}.  It is
derived in Sect.~3.  From this function, all other cases can be derived
(Sections~4--8).

Combining the results, we find the regions in the $g,\l$ plane where the
planar diagram summations (over different kinds of planar diagrams),
$$\sum_{V_3,V_4} N_a(V_3,V_4,E)g^{V_3} \l^{V_4}\,,\eqno(1.3)$$ converge.
Here, $N_a,\ a=0,1,2,\dots$ refers to the number of diagrams with different
kinds of restrictions, labled by the number $a$.

An accurate picture of the resulting regions is shown in Fig.~7.
Conclusions are summed up in Sect.~10.

Many of our results were obtained and/or checked using the computer program
{\sl Mathematica}.

\secbreak
{\ni\bf 2. General notation}

\nc We define
$$\eqalign {V_3&\iz \#\  \hbox{3-point vertices}\,,\cr
	V_4&\iz\#\  \hbox{4-point vertices}\,,\cr
	E&\iz\#\ \hbox{external lines}\,.} \eqno(2.1)$$
For the generating function for the number of Green functions that {\it include the
disconnected diagrams}, we use the symbol $G$:
$$G(\h,g,\l)\equiv\sum_{ E=0,\atop V_3=0,\,V_4=0}^{\infty,\,\infty,\,\infty}
\h^E\,g^{V_3}\,\l^{V_4}\,G_{(E,V_3,V_4)}\ .\eqno(2.2)$$
Most often, we shall concentrate on the generating functions for {\it connected
diagrams\/} only. They will be denoted by the letter $F$:
$$F(z,g,\l)\equiv\sum_{ E=1,\atop V_3=0,\,V_4=0}^{\infty,\,\infty,\,\infty}
z^{E-1}\,g^{V_3}\,\l^{V_4}\,F_{(E-1,V_3,V_4)}\ .\eqno(2.3)$$
Here, it is for technical reasons that we start counting at $E=1$, which will become clear.
Note, also, that we use a different symbol $z$, instead of $\h$. This is also
for later convenience.

The {\it one-particle irreducible\/} (1PI) diagrams will be generated by
the function $\G$:
$$\G(u,g,\l)\equiv\sum_{ E=1,\atop V_3=0,\,V_4=0}^{\infty,\,\infty,\,\infty}
u^{E-1}\,g^{V_3}\,\l^{V_4}\,\G_{(E-1,V_3,V_4)}\ .\eqno(2.4)$$

Apart from limiting ourselves to connected or irreducible diagrams,
we can also make restrictions on the occurrence of insertions within the
(reducible or irreducible) regions of the diagrams. This we indicate by
using a subscript $a$, taking values between 0 and 5, defined as in
the following list. The symbols $G$, $F$, $\G$, and the variables
$g,\l,\h,z,$ and $u$, are all replaced, successively, by $G_a$, $F_a$, $\G_a$,
$g_a,\l_a,\h_a,z_a,$ and $u_a$, meaning:
$$\eqalignno{a=0,&\ (G_0,\,F_0,\ ,\dots):\quad\hbox{no further restrictions.}&(2.5)\cr
&a=1, \ (G_1,\,F_1,\ ,\dots):\quad\hbox{no tadpoles:}\epsffile{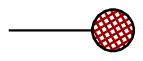}&(2.6)\cr
&a=2: \quad\hbox{no tadpoles and no seagulls:}\epsffile{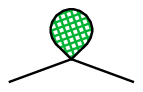}&(2.7)\cr
&a=3: \quad\hbox{also no self-energies:}\qquad \,\epsffile{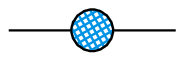}&(2.8)\cr
&a=4: \quad\hbox{also no dressed 3-vertices:}\epsffile{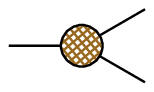}&(2.9)\cr
&a=5: \quad\hbox{also no dressed 4-vertices:}\epsffile{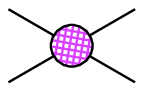}&(2.10)}$$ 
It will turn out to be important to use the subscripts also for the 
variables $g,\l,\h,z$, and $u$.
The reader may appreciate that the use of these suffixes is to be preferred
rather than exploiting the entire Latin and Greek alphabets. The price payed is
slightly more complicated-looking expressions. Where ambiguities can
be avoided, we will suppress some of the subscripts.

In Eq.~(1.1), our action is defined. Notice, that, in contrast with usual
conventions, we chose the signs in front of the interaction terms to be
positive. Therefore, the positive values of the coupling constants are the
unstable ones. The reason for this unorthodox choice is that now the expansions
in terms of $g_a$ and $\l_a$ all carry plus signs. With this sign convention, the
coefficients directly correspond to the number of diagrams. Furthermore,
unlike Ref\ref5, we chose factors $\fract13$ and $\fract14$ in front of the
coupling terms. This is quite conventional. In our case, these are the best combinatorial
factors to choose. The coefficients of all expansion terms will be integers,
and the only multiplicity factors with which diagrams are counted, are the ones
corresponding to how many ways a diagram can be rotated into distinct forms.
In Fig.~1, these multiplicity factors for some diagrams are indicated. 
\midinsert\cl{\epsffile{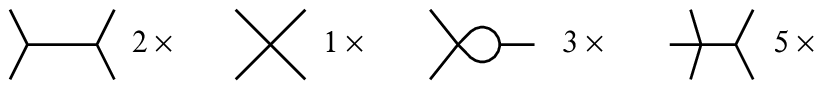}}
\cl{Figure 1.}
\cl{Multiplicity factors for some diagrams.}\endinsert
\secbreak

{\ni\bf 3. The primary equations}

\nc The recursion equations are most easily derived for the generating function $F=F_0$
describing all connected planar diagrams. Diagrammatically, we have 
Fig.~2. 
Since, in this entire section, we will
be limiting ourselves to the case $a=0$ (no further restrictions), we temporarily
omit the subscripts 0 that indicate this (so $F,\,g,\,\l,\,\dots$ stand for 
$F_0,\,g_0,\,\l_0,\,\dots$). 

\midinsert \cl{\epsffile{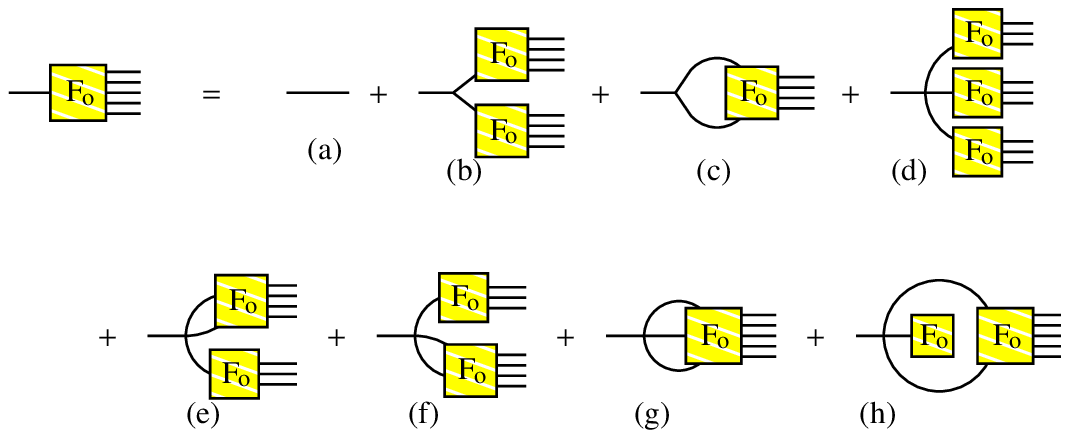}}
\cl{Figure 2. \ Equation (3.2)}\endinsert
Writing $$F(z,g,\l)=F_{(0)}(g,\l)+z\,F_{(1)}(g,\l)+z^2\,F_{(2)}(g,\l) +\dots 
\eqno(3.1)$$ 
(using brackets to distinguish the subscripts of the Taylor expansion from the
subscript $a$), the equation of Fig.~2 reads:
$$\eqalign{F=z&\ +\ g\Big(F^2\ +\ {F-F_{(0)}\over z}\Big)\ +\ \cr
	&\l\left(F^3\ +\  2F{ F-F_{(0)} \over z}\ +\ {F-F_{(0)}-F_{(1)}z\over z^2}
	\ +\  F_{(0)}{F-F_{(0)} \over z}\right)\,.}\eqno(3.2)$$
The last term is very important. If one does not take it into account, the resulting
equations, at a later stage, lead to horrendous complications impeding explicit
solution\fn{In principle, Eqs. of the form~(3.16)--(3.21) can then still be derived, but 
they become expressions 
involving polynomials whose degrees are well over 100, containing coefficients that 
are integers with hundreds of decimal places.}

The contributions of $F_{(0)}$ and $F_{(1)}$ are necessary for removing unwanted 
contributions from Green functions with too few external lines to close the loops 
in diagrams (c) and (e)--(h) in Fig.~2.

Clearly, one can solve Eq.~(3.2) if $F_{(0)}$ and $F_{(1)}$ are known. The trick to
find these was described by Koplik et al\ref3, and goes back to Tutte\ref4.
However, we need the result for general $g$ and $\l$, and it appears to be
difficult to extend their method directly. The matrix integration procedure used
by Br\'ezin et al\ref5, for pure $g\,M^3$ and pure $\l\,M^4$ theory, does work, also
for our more general case, but we can easily reproduce this general result
without integrating over matrices. What needs to be done, as Br\'ezin et al did, is
first to concentrate on the general Green functions $G_0$, instead of the
connected ones, $F_0$.

The relation between $G_a$ and $F_a$ is simple to derive diagrammatically. 
Consider a (connected or disconnected) diagram $G$. Draw a circle around it, such
that the circle is cut into $E$ pieces by the $E$ external lines. Select one of
these $E$ pieces as a starting point. This breaks the rotational symmetry, and
consequently any diagram with $E$ external lines will be counted $E/S$ times,
where $S$ is the dimension of the rotational symmetry group of the diagram
(cf Fig.~1). We open the circle at this point, and then find the relation between
$F$ and $G$ diagrammatically. See Fig.~3. In this Figure, the wiggled line
represents the circle that was opened up. The cross is the arbitrary point at the
boundary that was selected.
\midinsert\cl{\epsffile{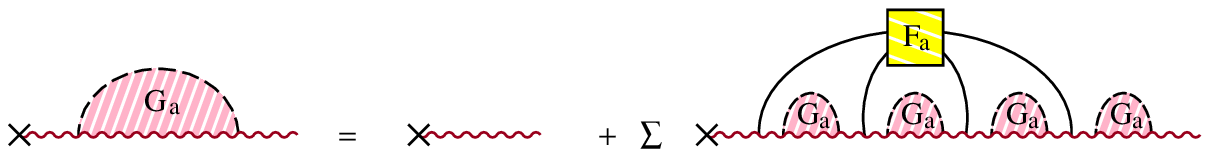}}
\cl{Figure 3. Recursive equation relating $G_a$ and $F_a$}
\endinsert
The resulting equation, which turns out to hold for all $a=0,\,\dots,\,5$, is
$$G(\h)=1+\sum_{E=1}^\infty F_{(E-1)}(g,\l)\big(G(\h)\big)^E\h^E=1+\h\,G(\h)F\big(
\h\, G(\h),g,\l\big)\,.\eqno(3.3)$$
Hence,
$$F\big(\h \,G(\h),g,\l\big)+{1\over \h\,G(\h)}={1\over\h}\,.\eqno(3.4)$$
Noticing, that $\h\,G(\h)$ acts as $z$ variable in the function $F$, we rewrite 
this equation by making the following identifications in Eqs.~(2.2) and (2.3):
$$\eqalignno{G_a&=1+z_a\,F_a={z_a\over\h_a },&(3.5)\cr
{1\over\h_a}&=F_a+{1\over z_a}\,.&(3.6)}$$
(valid for all $a=0,\dots,5$).

Substituting (3.6) into (3.2) gives us the recursive equation for $G(\h)$:
$$z+(x-{1\over z})(-1+gx+\l x^2)-{1\over z}(\a x+\b)=0\,,\eqno(3.7)$$
where
$$\eqalignno{x&={1\over\h}=F+{1\over z}\,;&(3.8)\cr
\a(g,\l)&=\l F_{0(0)}\ ;\qquad\b(g,\l)=g F_{0(0)}+\l F_{0(1)}+\l F_{0(0)}^2\,;&(3.9)\cr
z&=\h\,G(\h)\,,&(3.10)}$$ and all this only holds for $a=0$.
The symbol $x$ is the same as the one used to describe the Eigenvalues of the
matrices $M$, by Br\'ezin et al\ref5.

As this equation is only quadratic\fn{Here, the 
last term of Eq.~(3.2), diagram (h) in Fig.~2, was crucial.}  in $z$, it is easy to solve:
$$2z=x-gx^2-\l x^3\pm\sqrt{x^2(1-gx-\l x^2)^2+4\big(\l x^2+(g-\a)x+\b-1\big)}
\,.\eqno(3.11)$$
The, as yet unkwown, functions $\a(g,\l)$ and $\b(g,\l)$ can now be determined
by repeating the arguments by Tutte\ref4 and Koplik et al\ref3, and Br\'ezin et al\ref5.
We contemplate the branch points by looking at the roots of the 
6th degree polynomial in $x$ under the square root symbol, as $\l,g,\a$ and $\b$
become small.
At $x$ small compared to $1/g$ and $1/\sqrt\l$ (which means $\h$ large compared to $g$
and $\sqrt\l$)), there are only two roots,
$$x\ =\ x_{1,2}\ \approx\ \pm2+\OO(g,\sqrt\l)\,.\eqno(3.12)$$
All other roots must be at much larger values of $x$, i.e. $\h=\OO(g,\sqrt\l)$.
Branch points at such small values of $\h$ are inadmissible, however, and
therefore they must cancel out pairwise. Thus, for the other roots, one must
demand
$$x_3=x_4\ ;\qquad x_5=x_6\ .\eqno(3.13)$$
This means that it must be possible to write Eq. (3.11) as
$$2z=x-g x^2-\l x^3\pm\l(x-x_3)(x-x_5)\sqrt{(x-x_1)(x-x_2)}\,.\eqno(3.14)$$
It can also be seen from Eq.~(3.5) that, for large $x$, the function $z$, with the appropriate
sign choice for the square root, goes as $1/x$, but we do not need this information
here.

\def\onder{\vphantom{\sum_(}}

In order to bring the ensuing expressions into a slightly simpler form, we
rewrite $$x_1+x_2=2p\ ;\qquad x_1 x_2=-4+p^2-4q\ .\eqno(3.15)$$ In view of
Eq.~(3.12), this implies that $p$ and $q$ will be of order $g$ and $\l$.
\def\indik{{\,}\hskip-20pt} The requirement that Eq.~(3.11) can be written in
the form (3.14), because of Eqs.~(3.13), implies two constraints, which fix
the functions $\a(g,\l)$ and $\b(g,\l)$.  Explicitly, what one finds is:
$$\eqalignno{g&\ =\ {p\,(3+3\,p ^2-3\,q ^2+2\,p
^2q)\over(1+q)(6-3\,{p^2}+{\,p ^4}
+12\,q-3\,{p^2}\,q+6\,{q^2})\onder}\,;&(3.16)\cr \l&\ =\ {2q-2\,p ^2+2\,q
^2-\,p
^2q\over(1+q)(6-3\,{p^2}+{p^4}+12\,q-3\,{p^2}\,q+6\,{q^2})\onder}\,;&(3.17)\cr \a&\
=\ {p\,(2\,q -2\,p ^2+2\,q ^2-p^2q)(3+p^4+3\,q +p^2\,q -3\,q ^2+p^2\,q
^2-3\,q ^3)\over
(1+q)(6-3\,{p^2}+{p^4}+12\,q-3\,{p^2}\,q+6\,{q^2})^2\onder}\,;&(3.18)\cr \b&\ =\
{\Big({\dys 12\,q -3\,p ^2+11\,p ^4-5\,p ^6+p ^8-28\,p ^2q+30\,p ^4\,q -6\,p
^6\,q\, +\atop \dys +\,32q^2-46\,p ^2q^2+19\,p ^4\,q ^2+24\,q ^3-20\,p ^2\,q
^3+\,p ^2\,q ^4-4\,q ^5}\Big)\over
(6-3\,{p^2}+{p^4}+12\,q-3\,{p^2}\,q+6\,{q^2})^2}\,,\ &(3.19)}$$
and with Eqs.~(3.9):
$$\eqalignno{ \indik A\equiv F_{0(0)}&\ =\ {p\, ( 3 + {p^4} + 3\,q
+ {p^2}\,q - 3\,{q^2} + {p^2}\, {q^2} - 3\,{q^3} ) \over 6 - 3\,{p^2} + {p^4}
+ 12\,q - 3\,{p^2}\,q + 6\,{q^2}\onder }\,; &(3.20)\cr \indik B\equiv F_{0(1)}&\ =\
{( 1 + q ) \, \pmatrix{ 36 - 33\,{p^2} + 33\,{p^4} - 13\,{p^6} + {p^8} +
132\,q - 105\,{p^2}\,q \cr+\,85\,{p^4}\,q - 18\,{p^6}\,q + 168\,{q^2} -
126\,{p^2}\,{q^2} + 77\,{p^4}\,{q^2} \cr -\, 6\,{p^6}\,{q^2} + 72\,{q^3} -
78\,{p^2}\,{q^3} + 31\,{p^4}\,{q^3} - {p^6}\,{q^3}\cr -\, 12\,{q^4} -
33\,{p^2}\,{q^4} + 6\,{p^4}\,{q^4} - 12\,{q^5} - 9\,{p^2}\,{q^5}} \over
(6-3\,{p^2}+{p^4}+12\,q-3\,{p^2}\,q+6\,{q^2})^2 }\,.&(3.21) }$$ 
The way to interpret these equations is that $p$ and $q$ should be eliminated, to
find $A$ and $B$ as functions of $g$ and $\l$. 

The elimination process induces singularities as branch points in the
4-dimensional space of complex $g,\l$ values.  These in turn determine the
convergence of the combined $(g,\l)$ expansion.  Since all expansion
parameters for $F_{0(0)}$ and $F_{0(1)}$ must be non-negative integers, the
singularities at real and positive values for $g$ and $\l$ are of particular
relevance.  They will further be discussed in Section~9 and Appendix C.

Two special cases are: 
\nc 1) $g\f^3$ theory:
$$ \l=0\ ;\qquad p^2={2q(1+q)\over 2+q}\ .\eqno(3.22)$$
In this case, direct solution\fn{Different procedures lead to different
auxiliary parameters\ref{3,4,5}, but these can all be related one to another by
one-to-one mappings. Up to a sign, the parameter $\s$ shown here is
the one used by Br\'ezin et al\ref5.} leads to a single parameter, 
$\s=g/p$, replacing $p$ and $q$:
$$\eqalignno{p^2&={2\s\over(1-\s)(1-2\s)}\ ;\qquad q \ =\  
 {2\s\over1-2\s}\ ;\qquad\a=0\ ; &(3.23)\cr
g^2&={\s^2\over p^2}\ =\   \half\s(1-\s)(1-2\s)\ ;&(3.24)\cr
\b\ &=\   g\,F_{0(0)}\ =\   {\s(1-3\s)\over 2(1-2\s)}\ .&(3.25)}$$
The last two equations determine the function $F_{0(0)}(g)$, which is
here the only function needed to solve Eq.~(3.2).

\nc 2) $\l\f^4$ theory:
$$\eqalignno{g\ =\   0\ ;&\qquad p\ =\   0\ ;\qquad\a=0\ ;&(3.26)\cr
\l\ =\   {q\over 3(1+q)^2}\ ;&\qquad F_{0(0)}\ =\   0\ ;&(3.27)\cr
\b\ =\   \l F_{0(1)}&\ =\   {q\,(3-q)\over 9\,(1+q)}\ .&(3.28)}$$
\secbreak
{\ni\bf 4. One-particle-irreducible diagrams} 
\nc The generating function $\G$ for one-particle-irreducible (1PI) diagrams
has been treated in Ref\ref3. Here, we briefly review the procedure.
The treatments of this Section hold for all cases $a=0,\dots,5$,
so we suppress the suffix $a$.

Let
$$\eqalign{\G(u)&=-u+g\,u^2+\l\,u^3+\hbox{loop corrections}\cr
&= -u+\G_{(0)}+\tl\G_{(1)}u+\G_{(2)}u^2+\,\dots\,,}\eqno(4.1)$$
where $\tl\G_{(1)}=\G_{(1)}+1$ denotes the self-energy contributions
containing at least one vertex. The relation between $\G$ and $F$
can be seen diagrammatically, see Fig.~4.
\midinsert\cl{\epsffile{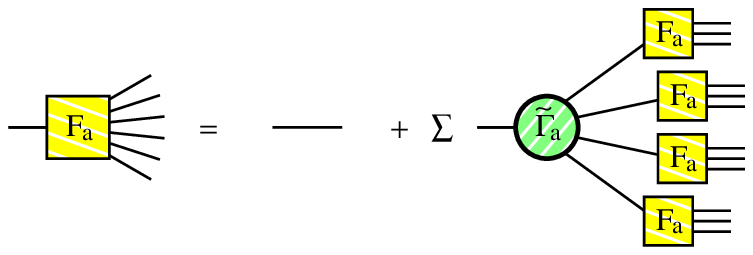}}
\cl{Figure 4. Equation (4.2) for $\G$}
\endinsert
This Figure corresponds to the equation
$$F(z)=z+\G_{(0)}+\tl \G_{(1)}\,F(z) +\G_{(2)}\,\big(F(z)\big)^2+\,\dots\,,
\eqno(4.2)$$
or:
$$-z=\G\big(F(z)\big)\,.\eqno(4.3)$$
We write
$$\eqalignno{\G_a&\ =\  -z_a\ =\  -\h_a\,G_a\,;&(4.4)\cr
u_a&\ =\  F_a\ =\  {1\over\h_a}-{1\over z_a}\ =\  {G_a-1\over z_a}\,.&(4.5)}$$

For a small number of external lines, and if $F_{(0)}$ and $\G_{(0)}$
vanish, the relations between the coefficients $\G_{(i)}$ and $F_{(i)}$
are straightforward:
$$ -z=\G_{(1)}\big(F_{(1)}z+F_{(2)}z^2+F_{(3)}z^3\dots\big)
+\G_{(2)}\big(F_{(1)}z+F_{(2)}z^2\dots\big)^2+\G_{(3)}\big(F_{(1)}z\dots\big)^3
+\dots\,,\eqno(4.6)$$ or
$$\eqalign{\G_{(1)}&=-{1\over F_{(1)}}\ ;\qquad \G_{(2)}={F_{(2)}\over F_{(1)}^2}
\ ;\qquad \G_{(3)}={F_{(1)}F_{(3)}-2\,F_{(2)}^2\over F_{(1)}^5}\ ;\cr
\G_{(4)}&={1\over F_{(1)}^7}\big[F_{(1)}^2F_{(4)}+5\,F_{(2)}^3-5\,F_{(1)}F_{(2)}
F_{(3)}\big]\ ,\qquad\hbox{etc.}}\eqno(4.7)$$
\secbreak
{\ni\bf 5. Removing tadpole insertions ($a=1$)}

\cl{\epsffile{gthplan1.ps}}
\nc Vacuum expectation values of single field operators often vanish.
In QCD, global gauge-invariance is sufficient to see that
$$ \bra 0| {A_\m}^i_j(x)|0\ket =0\,.\eqno(5.1)$$
In our case, this amplitude is $F(0,g,\l)=F_{0(0)}$. We now concentrate on the
case $a=1$, where we impose that all contributions of this sort are omitted.
Clearly, the last diagram (h) of Fig.~2 is now absent. But also diagrams (c)
and (g) require modifications, as they might produce the unwanted tadpole 
contributions. We thus obtain
$$F_{1(0)}=0\,,\eqno(5.2)$$ 
and removing all contributions to $F_{1(0)}$, we see that Eq.~(3.2) is replaced
by $$\eqalign{F(z,g,\l)&\ =\ z+ g\,\Big(F^2+{F-F_{(1)}z\over z}\Big)\ +\cr
&+\ \l\,\Big(F^3+{2F^2\over z}+{F-F_{(1)}z-F_{(2)}z^2\over z^2}\Big)\ ,\qquad
\hbox{if}\qquad a=1\,.}\eqno(5.3)$$
Here, $F,z,g,\l,\dots$ all stand for $F_1,z_1,g_1,\l_1,\dots$ 

There are two ways to proceed with this equation. First, one can again
write $x_1=F_1+1/z_1$, to obtain a quadratic equation for $z_1$ as a function
of $x_1$:
$$z_1+\big(x_1-{1\over z_1}\big)(\l_1 x_1^2+g_1 x_1-1)-\big(g_1 F_{1(1)}+\l_1 F_{1(2)}\big)
-{\l_1 F_{1(1)}\over z}\,,\eqno(5.4)$$ and we can proceed as in Sect.~3.

It is instructive, however, to find the direct relations between $F_1,\,z_1,\,g_1,\dots$
on the one hand, and $F_0,\,z_0,\,g_0,\dots$ on the other:
$$\eqalignno{F_1&=Q_1(F_0-F_{0(0)})\ ;&(5.5)\cr
z_1&={z_0\over Q_1}\ ,\qquad\hbox{hence}\qquad  x_1=Q_1(x_0-F_{0(0)})\ ;&(5.6)\cr
g_1&={g_0+3\l_0 F_{0(0)}\over Q_1^3}\ ;\qquad \l_1={\l_0\over Q_1^4}\ .&(5.7)}$$
The form of the function $Q_1(g,\l)$ can be read off from the requirement that, 
after the transformation, the Lagrangian again has a kinetic part normalized to one:
$$\LL\ =\ -\half x_0^2-\ffract13 g_0 x_0^3-\ffract14 \l_0 x_0^4\ =\ 
C+D x_1-\half x_1^2-\ffract13 g_1 x_1^3-\ffract14 \l_1 x_1^4\ ,\eqno(5.8)$$
from which one reads off Eqs.~(5.5)--(5.7), and
$$Q_1^2=1-2g_0F_{0(0)}-3\l_0 F_{0(0)}^2\,.\eqno(5.9)$$
The coefficient $D$ is adjusted so as to obey Eq.~(5.2), and the constant $C$
is irrelevant.
 
Indeed, one checks that with the substitutions (5.5)--(5.7) and (5.9), 
the Eq.~(3.2) turns into Eq.~(5.3).
In practice, this is the safe method: to adjust the transformations in such a way 
that the new recursion equation holds.

Since the coupling constants $g_1$ and $\l_1$ are now new functions of $p$ and $q$, 
their critical values will also be displaced. The fact that the critical values also 
follow the transformation rules (5.5)--(5.7) is not quite self-evident. But it is true,
and related to the general feature that $\dd F_{0(0)}$
and $\dd F_{0(1)} $ vanish as soon as $\dd p$ and $\dd q$ are such that 
$\dd g_0$ and $\dd \l_0 $ vanish.
We return to this issue in Sect.~9. 
\secbreak
{\ni\bf 6. Removing seagull insertions ($a=2$)}

\cl{\epsffile{gthplan2.ps}}
\nc The recursion relation for diagrams where tadpole and seagull insertions are 
removed, reads:
$$\eqalign{F(z,g,\l)&\ =\ z\ +\ g\,\big(F^2+{F-F_{(1)}z\over z}\big)\ +\cr
+&\ \l\,\Big(F^3\ +\ 2F{F-F_{(1)}z\over z}\ +\ {F-F_{(1)}z-F_{(2)}z^2\over z^2}\Big)\ ,}
\eqno(6.1)$$
where now $F,z,g,\l\,\dots$ stand for $F_2,z_2,g_2,\l_2,\,\dots$

As the seagulls only affect the dressed propagators, the relation between $F_2$
and $F_0$ is as in Eqs.~(5.5)--(5.7), but with a different choice for $Q$. We
find that Eq.~(6.1)  is obeyed provided that
$$\eqalignno{F_2&=Q_2(F_0-F_{0(0)})\ ;&(6.2)\cr
z_2&={z_0\over Q_2}\ ,\qquad\hbox{so}\qquad  x_2=Q_2(x_0-F_{0(0)})\ ;&(6.3)\cr
g_2&={g_0+3\l_0 F_{0(0)}\over {Q_2}^3}\ ;\qquad \l_1={\l_0\over {Q_2}^4}\ ;&(6.4) \cr
{Q_2}^2&=1-2g_0F_{0(0)}-3\l_0 F_{0(0)}^2-2\l_0 F_{0(1)}\ .&(6.5)}$$
 
\secbreak
{\ni\bf 7. Removing self-energy insertions ($a=3$)}

\cl{\epsffile{gthplan3.ps}}
\nc After having removed the tadpoles and the seagulls, we remove the self-energy
insertions. The recursion relation becomes, in diagrams, Fig.~5. The equation 
for the case $a=3$ becomes
$$F\ =\ z\ +\ g\Big(F^3-{F-1\over z}\Big)\ +\ \l\Big(F^3+{2F^2\over z}
+{F-z-z^2F_{(2)}\over z^2}\Big)\ -\ C\,F\,,\eqno(7.1)$$ with
$$F_{(1)}=1\ ,\qquad\hbox{or}\quad F=z+F_{(2)}z^2+\OO(z^3)\ .\eqno(7.2)$$

\midinsert\cl{\epsffile{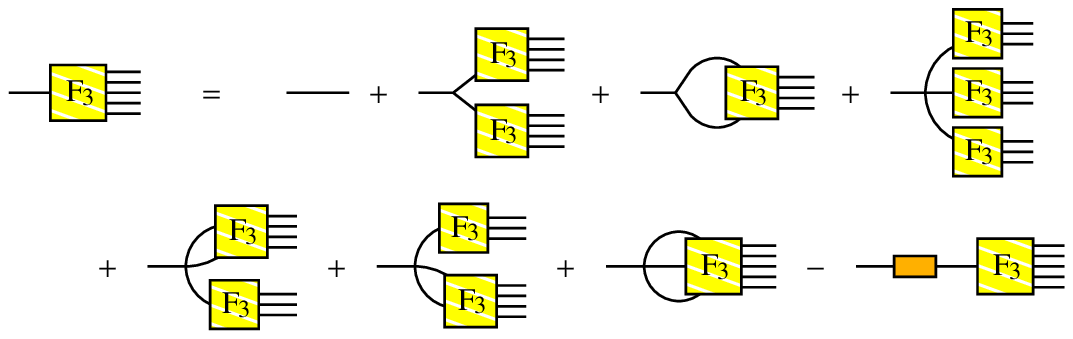}} \cl{Figure 5.  Removing
self-energies.}\endinsert 
In (7.1), the last term removes everything that
might have produced a self-energy insertion up front.  This includes the
seagull contribution, which is therefore removed automatically, so that the
corresponding term in Eq.~(6.1), being $-2\l\,F\,F_{(1)}$, could be left out.
It just readjusts the $z$-independent number $C$.

$C$ is now fixed by the requirement (7.2), $F_{3(1)}=1$. Comparing Eq.~(7.1) with
(5.3) fives us the relations
$$\eqalignno{F_3&=Q_3F_1\ ;\qquad z_3=z_1/Q_3\ ;\qquad 1+C=1/{Q_3}^2\ ;&(7.3)\cr
&g_3=g_1/{Q_3}^3\ ;\qquad \l_3 =\l_1/{Q_3}^4\ ; &(7.4)\cr
&{Q_3}^2=1/F_{1(1)}\ ;\qquad \ (Q_1Q_3)^2=1/F_{0(1)}\ .&(7.5)}$$
It is easy to see that these are merely field renormalizations that replace the 
dressed propagators by bare ones.
\secbreak
{\ni\bf 8. Removing dressed 3-vertices and dressed 4-vertices}

\cl{\epsffile{gthplan4.ps}\qquad\epsffile{gthplan5.ps}} \nc The last two
steps are remarkably easy.  Recursion relations for the last two cases are
not so easy to write down, but they are not needed.  The amplitudes
considered remain the same, but only the definitions of the coupling
constants change.  All one does is recombine all contributions from subgraphs
that express dressed 3-point vertex insertions and use these to redefine the
coupling constant $g$.  After this, we do the same with the 4-vertex
contributions to redefine $\l$.  See Fig.~6.
\midinsert\cl{\epsffile{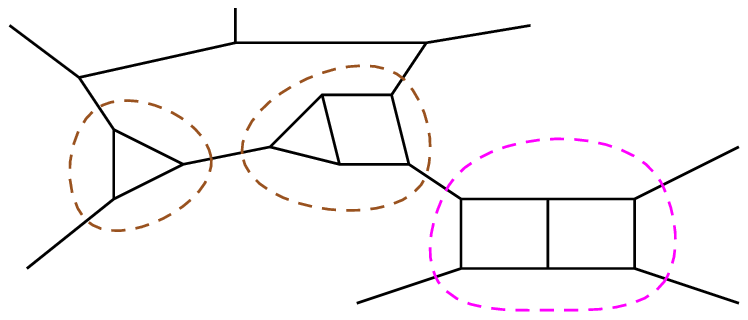}}

\cl{Figure 6. Collecting 3- and 4-point subgraphs.}\endinsert
This allows us to write:
$$\eqalignno{F_4&\ =\   F_3\ ;\qquad z_4\ =\ z_3\ ;&(8.1)\cr
g_4&\ =\ F_{3(2)}\ =\ {Q_3}^3F_{1(2)}\ =\ (Q_1Q_3)^3F_{0(2)}\ =\ {F_{0(2)}\over
{F_{0(1)}}^{3/2}}\ .&(8.2)\cr
\l_4&\ =\ \l_3\ ,&(8.3)}$$
and when we also remove the dressed 4-vertices:
$$\eqalignno{F_5&\ =\ F_3\ ;\qquad z_5\ =\ z_3\ ;&(8.4)\cr
g_5&\ =\ g_4\ =\ F_{0(2)}/{F_{0(1)}}^{3/2}\ ;&(8.5)\cr
\l_5&\ =\ \G_{3(1)}\ =\ F_{3(3)}-2\,{F_{3(2)}}^2\ =\ {F_{1(1)}F_{1(3)}-2{F_{1(2)}}^2
\over {F_{1(1)}}^3}&\cr
&\qquad\qquad\ =\ {F_{0(1)}F_{0(3)}-2\,{F_{0(2)}}^2\over {F_{0(1)}}^3}\ .
&(8.6)}$$
\secbreak
{\ni\bf 9. Critical couplings}
\nc At high orders, the number of diagrams can be read off from the values of $g$ and $\l$ where the 
amplitudes develop their first singularities. These occur when the elimination process
of $p$ and $q$ in Eqs.~(3.16)--(3.19) develops branch points. A branch point is a
set of values for $p$ and $q$ where a small variation $(\dd p,\,\dd q)$ produces
a vanishing variation $(\dd g,\,\dd\l)$. Or,
$${\rm det}\left|\matrix{\part g{p}\onder&\part gq\cr\part\l p&\part\l q}\right|=0\ .
\eqno(9.1)$$
The solution of this equation was found to be
$$\matrix{p\onder=t\sqrt z\ ;\cr q=z-1\ ;}\qquad z={12-4\,t^3+t^4\over 6-6\,t
-3\,t^2+2\,t^3}\ .\eqno(9.2)$$
At these points, we have $(g_a,\l_a)=(g_a^c,\l_a^c)$, with
$$\eqalignno{g_0^c&={\pm2\,(3-t)t^2\,(6-6\,t-3\,t^2+2\,t^3)^{1/2}\over
(12-4\,t^3+t^4)^{3/2}\onder}\ ;&(9.3)\cr
\l_0^c&={(2+2\,t-t^2)(6-6\,t-3\,t^2+2\,t^3)\over(12-4\,t^3+t^4)^2}\ .
&(9.4)}$$

The most important region, where $g>0$ and $\l>0$, is from $t=1-\sqrt3$
(where $\l_0^c=0,\,g_0^c=1/\sqrt{12\sqrt3}$) to $t=0$ (where $\l_0^c=1/12,\,
g_0^c=0$).

An important non-trivial finding is that for the critical line near the origin,
where Eq.~(9.1) holds, one also has, for the functions $\a$ and $\b$ defined in
Eq.~(3.9),
$${\part\a p\Big/\part\a q}={\part\b p\Big/ \part\b q}={\part gp\Big/\part gq}
={\part\l p\Big/\part\l q}\,,\eqno(9.5)$$
so that, at a critical point where $(\dd p,\,\dd q)\ne0$, but $(\dd g,\,\dd\l)=0$,
one also has $(\dd\a,\,\dd\b)=0$, and since all other coupling parameters
$g_a,\,\l_a$ are built from $\a$ and $\b$, we may conclude that the critical
values of these other coupling constants directly follow from the critical
values of the parameters $p$ and $q$. In other words, Eqs.~(9.2) apply to
all critical couplings $g_a,\l_a$, $a=0,\dots,5$. Although this is what one
would have expected on physical grounds (the divergence of the diagrams should not
depend on the choice of coupling parameters), the mathematical reason for
this phenomenon is not totally clear to the author.

The results for all critical lines are deferred to Appendix C.
We depict the critical lines in Fig.~7. We see the regions of convergence grow as
more constraints are imposed on the diagrams. The most surprising feature perhaps 
is the kinks in the curves at $g=0$ (since diagrams assemble in even or odd functions
of $g$, the picture is symmetric). One would have expected smooth, tilted lines
if $\l$ is plotted against $g^2$, not $g$, so that the lines should run
horizontally at $g=0$ in the $g\,\l$ plane. But such is not the case. 
The complete courses in the ($g^2, \l$) plane of the critical curves at all $t$
values is quite complex, showing cusps, among others at $g^2=0$. In contrast,
only the physical regions shown in Fig.~7 appear to be rather featureless.
(the figure is highly accurate).

\midinsert\cl{\epsffile{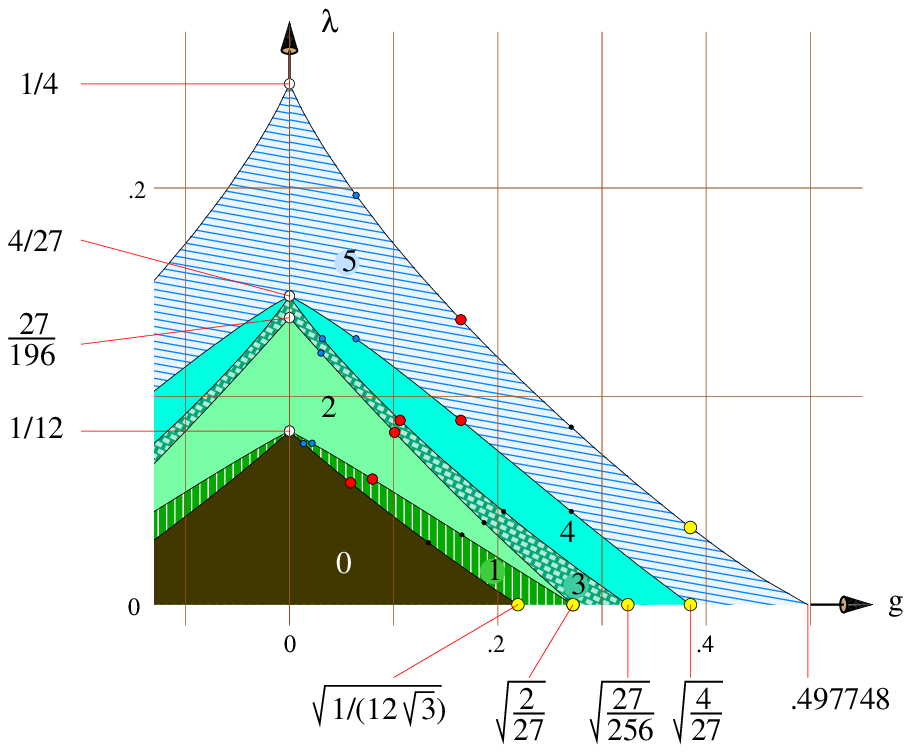}}

\cl{Figure 7. Convergence regions in the $g,\,\l$ plane}
\Narrower The convergence domains for the cases $a=0$ to $5$ are shown. The picture
is numerically accurate. Dots of equal colors and sizes correspond to
equal values of $p$, $q$ and $t$.\Endnarrower\endinsert 

At zero $g$ and at zero $\l$, the critical points are simple algebraic numbers,
as indicated in Fig.~7; only at $\l^c_5=0$, the value of $g^c_5$ is a root of a
polynomial of a high degree.
\secbreak
{\ni\bf 10. Conclusion}
\nc All interesting sets of planar diagrams can be counted, in the sense that the
generating functions for their numbers can be found analytically. This
provides us with exact descriptions of the critical lines in the $g,\,\l$ plane.
These lines show kinks at $g=0$, where one would have expected analytic
$g^2$ dependence. Fig.~7 shows the convergence regions accurately, when
$g$ and $\l$ are positive and real. The detailed mathematical expressions 
are given in the appendices B and C. 
\secbreak
{\ni\bf References}
\item{1.} G.~'t Hooft, Nucl. Phys. {\bf B72} (1974) 461; Nucl. Phys. 
{\bf B75} (1974) 461.  
\item{2.} H.B.~Nielsen and P.~Olesen, Phys.~Letters {\bf 32 B} (1970)203;\br
B.~Sakita and M.A.~Virasoro, Phys.~Rev.~Letters {\bf 24} (1970) 1146;\br
H.~Bohr and H.B.~Nielsen, Nucl.~Phys. {\bf B227} (1983) 547. 
\item{3.} J.~Koplik, A.~Neveu and S.~Nussinov, Nucl.~Phys. {\bf B 123} (1977)109
\item{4.} W.T.~Tutte, Can. J. Math. {\bf 14} (1962) 21.
\item{5.} E.~Br\'ezin, C.~Itzykson, G.~Parisi and J.B.~Zuber, Commun.~Math.~Phys.
{\bf 59} (1978) 35. 
\item{6.} G.~'t Hooft, Commun.~Math.~Phys. {\bf 88} (1983) 1; "Planar diagram
field theories", in {\it Progress in Gauge Field Theory, NATO Adv.~Study Inst.~
Series}, eds.~G.~'t Hooft et al., Plenum, 1984, p. 271.
\secbreak
\nc
{\ni\bf Appendix A. Constructing perturbative planar QCD amplitudes at all orders
without divergences} 

\nc A renormalization scheme that appears to be not so well-known\ref6 can be
set up in the following way.  It works particularly well for planar field
theories in 4 space-time dimensions, but it also applies to several other
quantum field theories, notably QED.  Presumably, QCD with $N_c=3$ can also
be covered along hese lines, but some technical details have not been worked
out to my knowledge. This appendix gives a brief description.

Our scheme consists of first collecting all one-particle irreducible 2-, 3-
and 4-point diagrams, and formally considering the non-local quartic effective
action generated by these diagrams.

Next, consider all diagrams using the Feynman rules derived from this action,
but {\it with the limitation that only those diagrams that are absolutely
ultraviolet vonvergent are included}.  No superficially ultraviolet divergent
subgraph is accepted.  To be precise, we omit all diagrams with 4 or less
external lines, as well as all diagrams containing any non-trivial subdiagram
with 4 or less external lines.

Clearly, one expects that the 1PI subgraphs with 4 or less external lines
have already been taken care of by our use of the quartic effective action
instead of the bare Lagrangian.  The important issue to address is, whether
the counting of all diagrams was done correctly so as to onbtain the required
physical amplitudes.  But this is not difficult to prove:

$\underline{\hbox{Theorem:}}$ the above procedure correctly reproduces the
complete amplitudes for the original theory.  No diagrams are over-counted or
under-counted.

The proof of the theorem is by inspection.  Take any diagram of the original
theory.  Consider all its 2-, 3-, and 4-point subgraphs, and ascertain that
these can be identified unambiguously as contributions of our effective
Lagrangian.  In practice, one draws circles around the subgraphs as
illustrated in Fig.~6.  Over- or under-counting can only happen if any
ambiguity would arise with the identification.  Such ambiguities can only be
expected if two subgraphs are connected by at least two lines.  But if they
both would have 4 external lines or less, the entire (sub-)diagram would have
$4+4-2\times2=4$ or less external lines, and this implies that this entire
combination should itself be counted as one single contribution to the
effective Lagrangian (see Fig.~8a).

If we would have tried to continue the procedure by including 5-point irreducible
subgraphs into a quintic effective Lagrangian (or more), then counting
problems would arise: $5+5-2\times2>5$ (see Fig.~8b)
\midinsert\cl{\epsffile{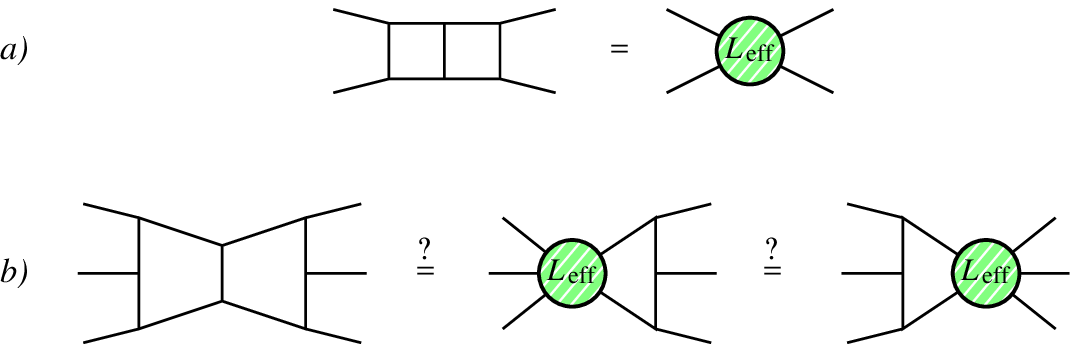}}
\Narrower Figure 8. Ambiguity when including a 5-point vertex in the effective
Lagrangian\Endnarrower\endinsert

From our theorem, one deduces that\br
{\it if all one-particle irreducible 2-, 3-, and 4-point vertices are known,
all amplitudes can be derived without encountering any divergent
(sub-)graph(s).}\br
In a planar theory, the diagrams considered are of our class~5. 

What remains to be done to complete a perturbative computational scheme, is
to establish an algorithm to compute the irreducible 2-, 3-, and 4-point
vertices (and, if they occur, the tadpole diagrams as well). Actually, this
is simple. Consider a 4-point 1PI diagram $\G(p_1,p_2,p_3,p_4)$. Here,
$p_\m$ are the external momenta, and $p_1+p_2+p_3+p_4=0$. Now consider
the difference
$$\G(p_1+k,p_2-k,p_3,p_4)-\G(p_1,p_2,p_3,p_4)\equiv k_\m\D^\m(p_1,
\underline{k},p_2-k,p_3,p_4) \,.\eqno(A.1)$$ 
The underlining refers to the fact that, in the function $\D^\m$,
this external line follows distinct Feynman rules. 

If we follow a path inside the diagram, we can consider the entire expression 
for $\D^\m$ as being built from expressions containing differences. For instance,
in the propagators:
$${1\over(p+k)^2+m^2}-{1\over p^2+m^2}\ =\ {k_\m\,(-2p-k)^\m\over
\big((p+k)^2+m^2\big)(p^2+m^2)}\,,\eqno(A.2)$$
or else, in the 3-vertices:
$$(p+k)_\n\ -\ p_\n\ =\ k_\m{\d^\m}_{\n}\ .\eqno(A.3)$$
We notice, that the expressions for $\D^\m$ are all (superficially) 
ultraviolet convergent! Actually, one may set up unambiguous Feynman rules 
for $\D^\m(p_1, \underline{k},p_2,-kp_3,p_4)$
and observe that this amplitude exactly behaves as a 5-point diagram, hence it is
(superficially) convergent.

For the 3-point and the 2-point diagrams, one can do exactly the same
thing by differentiating more than once. In practice, what one finds is,
that there is a set of rules containing fundamental irreducible 2-, 3- and 4-point
vertices, and in addition rules to determine their differences at different
values of their momenta. The complete procedure thus leads to the following
situation. 

\def\uunderl#1{\underline{\underline{#1}}}

We start by postulating the so-called `primary vertex functions'. These are,
not only the irreducible 2-point functions\fn{Here, the entries in the square 
brackets $[\dots]$ differ from the ones in the curved brackets $(\dots)$ of
Sect.~(4) by one unit.} $\G_{[2]}(p,-p)$, the irreducible 3-point functions
$\G_{[3]}(p_1,p_2,-p_1-p_2)$ and the irreducible 4-point functions
$\G_{[4]}(p_1,\dots,p_4)$, but also, in addition, the difference functions
$\D^\m_{[2]}(p,\underline{k},-p-k)$ and $\D^\m_{[3]}(p_1,\underline{k}
,p_2-k,-p_1-p_2)$, and
finally the functions $U_{[2]}$, obtained by differentiating $\D^\m_{[2]}$ once more:
$$\D_{[2]}^\m(p_1+q,p_2-q,-p_1-p_2)-\D_{[2]}^\m(p_1,p_2,-p_1-p_2)\equiv
q_\n U_{[2]}^{\m\n}(p_1-q,\uunderl{q},p_2,-p_1-p_2)\,,\eqno(A.4)$$
where one of the other external lines, $p_1$ or $p_2$ is underlined.
The double underlining is here to denote that the two entries are to
be treated distinctly (because of the factor $k_\m$, the functions $U_{[2]}^{\m\n}$
are not symmetric under interchange of $k$ and $q$).

These primary vertex functions are derived by first considering the differences 
for $\G_{[4]}$, $\D^\m_{[3]}$ and $U^{\m\n}_{[2]}$ at two different sets
of external momenta. These expressions are handled as if they were irreducible
5-point diagrams. These are expanded in terms of planar diagrams where all irreducible
subraphs of 4 or less external lines are bundled to form the primary vertex
functions. At one of the the edges of such a diagram, we then encounter one
of the functions $\D^\m$ or $U^{\m\n}$.

This way, we arrive at difference equations for the primary vertices, with at
the r.h.s. again the primary vertices. The primary vertices $\G_{[3]}$,
$\D^\m_{[2]}$ and $\G_{2]}$ are then obtained by integrating 
$\D^\m_{[3]}$, $U^{\m\n}_{[2]}$ and $\D^\m_{[2]}$ with respect to 
the external momenta. This completes the procedure to obtain all amplitudes
by iteration.

Technical implementation of our scheme requires that in all diagrams, an
unambiguous path can be defined from one external line to another. In QED, 
one may use the paths defined by the electron lines. In a planar theory, 
one may define the paths to run along the edges of a diagram.

Three remarks are of arder:
\item{1.} The procedure is effectively a renormalization group procedure.
The functions $\D^\m$ and $U^{\m\n}$ play the role of beta functions.
\item{2.} The procedure is essentially still perturbative, since the planar diagrams
must still be summed. Our beta functions are free of ultraviolet divergences,
but the summation over planar diagrams may well diverge.
\item{3.} The procedure only works if the integrations do not lead to
clashes. This implies that it is not to be viewed as a substitute for
regularization procedures such as dimensional regularization. We still need
dimensional regularization if we want to {\it prove\/} that the method is unambiguous,
which, of course, it is, in the case of planar QCD.
\secbreak
{\ni\bf Appendix B. The power expansions in $g$ and $\l$.}
\nc
To illustrate how our expressions count diagrams, we produce the first
few terms of the power expansions in $g$ and $\l$.

\def\frac#1#2{{#1\over#2}}

\ni{\it  The case $a=0$.}\br Let us consider the double series expansions
of $g$ and $\l$ with respect to $p$ and $q$.  Expanding Eqs.~(3.16) and
(3.17), we get $$\eqalign{ g_0 \,=\,& \left({\frac{1}{2}} -{\frac{3}{2}}\,q +
{\frac{5}{2}}\,q ^2 -{\frac{7}{2}}\,q ^3 +{\frac{9}{2}} \,q ^4
-{\frac{11}{2}}\,q ^5 + {\frac{13}{2}}\,q ^6 \right)\,p\,+\cr
&\left({\frac{3}{4}} -{\frac{13}{6}}\,q + {\frac{17}{4}} \,q ^2 -7 \,q ^3 +
{\frac{125}{12}}\,q ^4 -{\frac{29}{2}} \,q ^5 +{\frac{77}{4}}\,q ^6 \right)
\,p^3\, + \cr & \left( {\frac{7}{24}}-{\frac{25}{24}} \,q +{\frac{29}{12}}\,q
^2 -{\frac{55}{12}}\,q ^3 +{\frac{185}{24}}\,q ^4-{\frac{287}{24}}\,q
^5+{\frac{35}{2}} \,q ^6\right) \,p^5+ \OO(p^7,\,q ^7)\,;}\eqno(B.1)$$
$$\eqalign{ \l_0\,=\,& {\frac{1}{3}}\,q -{\frac{2}{3}}\,q ^2 +q^3
-{\frac{4}{3}}\,q ^4+ {\frac{5}{3}}\,q ^5 -2\, q^6 +\cr & \left(
-{\frac{1}{3}}+ q -2\,q^2+ {\frac{10}{3}}\,q ^3 -5\,q^4 + 7\,q^5
-{\frac{28}{3}}\,q ^6 \right) \,p^2+\cr &\left( -{\frac{1}{6}}+
{\frac{11}{18}}\,q -{\frac{13}{9}}\,q ^2+ {\frac{25}{9}}\,q ^3
-{\frac{85}{18}}\,q ^4 +{\frac{133}{18}}\,q ^5 -{\frac{98}{9}}\,q ^6
\right)\,p^4 +\OO(p^6,q^7)\,.}\eqno(B.2)$$

\ni Such expansions will be displayed in a short-hand notation:
$$g_0(p,q)=\quad\bordermatrix{&1&q&q^2&q^3&q^4&q^5&q^6\onder\cr
p&{\frac{1}{2}} & -{\frac{3}{2}} & {\frac{5}{2}} & -{\frac{7}{2}} &
{\frac{9}{2}} & -{\frac{11}{2}} & {\frac{13}{2}}\onder\cr p^3& {\frac{3}{4}}
& -{\frac{13}{6}} & {\frac{17}{4}} & -7 & {\frac{125}{12}} & -{\frac{29}{2}}
& {\frac{77}{4}}\onder\cr p^5& {\frac{7}{24}} & -{\frac{25}{24}} &
{\frac{29}{12}} & -{\frac{55}{12}} & {\frac{185}{24}} & -{\frac{287}{24}} &
{\frac{35}{2}}\cr}\eqno(B.3) $$
$$\l_0(p,q)=\quad\bordermatrix{&&&&&\cr &0 & {\frac{1}{3}} & -{\frac{2}{3}} &
1 & -{\frac{4}{3}} & {\frac{5}{3}} & -2\onder\cr p^2 & -{\frac{1}{3}} & 1 &
-2 & {\frac{10}{3}} & -5 & 7 & -{\frac{28}{3}} \onder \cr p^4 &
-{\frac{1}{6}} & {\frac{11}{18}} & -{\frac{13}{9}} & {\frac{25}{9}} &
-{\frac{85}{18}} & {\frac{133}{18}} & -{\frac{98}{9}} \cr}\eqno(B.4)$$
Of course, the matrices extend to infinity.  We invert the series\fn{For the
inversion, a few more terms not shown in Eqs.~(B.3--4) were needed.  Similarly,
Equs. (B.11) and (B.12) required a few terms not shown in (B.9,10). The
coefficients that we do show are the correct first few terms of the infinite series
expansions.} \ in $x$ and $y$:
$$\eqalignno{p(g_0,\l_0)&=\quad\bordermatrix{& &\l_0&\l_0^2&\l_0^3&\l_0^4&\l_0^5\cr g_0
& 2 & 18 & 180 & 1890 & 20412 & 224532\cr g_0^3& 12 & 368 & 7860 & 143424 &
2393496 & 37700640\cr g_0^5& 128 & 7120 & 240768 & 6390720 &146382336 &
3033374832\cr}&(B.5)\cr
 q(g_0,\l_0)&=\quad\bordermatrix{&&&&&\cr &0 & 3
& 18 & 135 & 1134 & 10206\cr g_0^2 & 4 & 84 & 1368 & 20196 & 283176 &
3847176\cr g_0^4 & 40 & 1768 & 49656 & 1127808 & 22584528 &
415844280\cr}&(B.6)}$$

\ni The first Green functions are then $$ \eqalignno{F_{0(0)}(g_0,\l_0)&=
\quad\bordermatrix{&&&&\cr g_0&1 & 6 & 45 & 378 & 3402 & 32076\cr g_0^3 & 4 &
92 & 1572 & 23904 & 341928 & 4712580\cr g_0^5& 32 & 1424 & 40128 & 912960 &
18297792 & 337041648\cr}&(B.7)\cr
F_{0(1)}(g_0 ,\l_0)&= \quad\bordermatrix{&&&&\cr &1 & 2 & 9 & 54 & 378 &
2916\cr g_0^2 & 3 & 48 & 630 & 7776 & 93555 & 1111968\cr g_0^4 & 24 & 856 &
20112 & 392040 & 6868152 & 112295160\cr}&(B.8)}$$

\ni{\it  The case $a=1$.}\br
From now on, we leave the powers of $g$ and $\l$ to be understood.  Whether
the $g$'s come in even or odd powers depends on whether the number of
external lines is even or odd.  Before looking at these numbers, we continue
to produce more lists. From Eqs.~(5.6) and (5.7):  
$$ \eqalignno{g_1(p,q)&=\quad\pmatrix{
{\frac{1}{2}}&-1&1&-1&1&-1\onder\cr
{\frac{5}{8}}&-{\frac{95}{48}}&{\frac{109}{24}}&-{\frac{65}{8}}&
{\frac{301}{24}}&-{\frac{427}{24}}\onder\cr  
{\frac{149}{192}}&-{\frac{823}{192}}&{\frac{10453}{768}}& 
-{\frac{4183}{128}}&{\frac{25513}{384}}&-{\frac{46309}{384}}\onder\cr}&(B.9)\cr  
 \l_1(p,q)&=\quad\pmatrix{0&{\frac{1}{3}}&-{\frac{2}{3}}&1&-{\frac{4}{3}}
 &{\frac{5}{3}}\onder\cr  
-{\frac{1}{3}}&{\frac{4}{3}}&-{\frac{23}{6}}&{\frac{26}{3}}& 
-{\frac{33}{2}}&28\onder\cr   -{\frac{1}{2}}&{\frac{127}{36}}&-{\frac{499}{36}}& 
{\frac{5765}{144}}&-{\frac{6883}{72}}&{\frac{28807}{144}} \cr}&(B.10)}$$ 
Inverting this gives:
$$\eqalignno{p(g_1,\l_1)&=\quad\pmatrix{2 & 12 & 108 & 1080 & 11340 & 122472\cr  6 & 143 & 2700 
& 45441 & 714528 & 10741086\cr  35 & 1685 & {\frac{201447}{4}} &  {\frac{2425941}{2}} 
& 25670925 & 497809800\onder \cr
}&(B.11)\cr
 q(g_1,\l_1)&=\quad\pmatrix{0 & 3 & 18 & 135 & 1134 & 10206\cr 4 & 48 & 666 & 9072 &
 120852 & 1582416\cr 
16 & 592 & 14328 & 292869 & 5416524 & 93623769\cr}&(B.12)}$$
Now, $F_{1(0)}=0$. The next Green functions are found by expanding Eq.~(5.5),
using (5.3) 0r (3.2):
$$\eqalignno{F_{1(1)}(g_1,\l_1)&=\quad\pmatrix{1 & 2 & 9 & 54 & 378 & 2916\cr 1 & 15 & 189 & 2268 & 
26730 & 312741\cr
 4 & 132 & 2925 & 54432 & 917973 & 14535288\onder\cr}&(B.13)\cr
 F_{1(2)}(g_1,\l_1)&=\quad\pmatrix{  1 & 9 & 81 & 756 & 7290 & 72171   \cr 4 & 99 & 1755 
 & 27216 & 393417 & 5450733 \cr  
 24 & 1044 & 28674 & 635607 & 12420216 & 223297074  \cr 
 }&(B.14)}$$

\ni{\it The case $a=2$. }\  Similarly, we found:
$$\eqalignno{ F_{2(1)}(g_2,\l_2)&=\quad\pmatrix{1 & 0 & 1 & 2 & 10 & 42\cr 1 & 5 & 35 & 228 & 1540 & 10439\cr
 4 & 60 & 725 & 7636 & 74725 & 695464\onder\cr}&(B.15)\cr
  F_{2(2)}(g_2, \l_2)&=\quad\pmatrix{1 & 3 & 15 & 76 & 420 & 2409\cr 4 & 45 & 435 & 
3818 & 32025 & 260799\cr
24 & 540 & 8370 & 107877 & 1245960 & 13365702\cr}&(B.16)}$$

\ni{\it The case $a=3$:} \ $F_{3(1)}=1$ and
$$F_{3(2)}(g_3,\l_3)=\quad\pmatrix{1 & 3 & 12 & 55 & 273 & 1428\cr 1 & 15 & 159 & 
1460 & 12405 & 100449\cr 
 3 & 90 & 1638 & 23400 & 288738 & 3227490\cr}\eqno(B.17)$$
Concentrating now on the 1PI functions: 
$$\eqalignno{\G_{3(3)}(g_3,\l_3)&=\quad\pmatrix{0&  1&  2&  6&  22&  91 \cr 0& 4& 44& 364& 2720& 
19380& \cr 1& 34&  596&  7852&  88251&  896972 \onder
\cr}&(B.18)\cr
\G_{3(4)}(g_3,\l_3)&=\quad\pmatrix{ 0&0&5&45&315&2040 \cr 0&5&115&1565&16950&161950 \cr  
1&65&1750&31890&465080&5873405\onder \cr} &(B.19)\cr
\G_{3(5)}(g_3,\l_3)&=\quad\pmatrix{ 0&0&0&2&15&90 \cr 0&0&9&168&1938&18240 \cr 
 0&6&249&5176&77841&970596  \cr}&(B.20)}$$

\ni{\it $a=4$:} \ $\G_{4(2)}=g_4$ and
$$\eqalignno{\G_{4(3)}(g_4,\l_4)&=\quad\pmatrix{  0&1&2&6&22&91   \cr 0&4&20&112&660&4004 \cr 
  1&14&142&1288&10990&90174 \onder\cr}&(B.21)\cr
  \G_{4(4)}(g_4,\l_4)&=\quad\pmatrix{  0&0&5&30&165&910   \cr 0&5&65&560&4365&32585 \cr 
  1&35&550&6580&68740&661647 \onder \cr}&(B.22)\cr
  \G_{4(5)}(g_4,\l_4)&=\quad\pmatrix{ 0&0&0&2&15&90 \cr 0&0&9&114&957&7098 \cr 
  0&6&159&2104&21909&203370 \cr}&(B.23)}$$

\ni{\it $a=5$:} \ $\G_{5(3)}=\l_5$ and
$$\eqalignno{\G_{5(4)}(g_5,\l_5)&=\quad\pmatrix{  0&0&5&10&25&70  \cr 0&5&15&70&355&1770 \cr 
 1&5&50&430&3240&22422 \onder \cr}&(B.24)\cr
 \G_{5(5)}(g_5,\l_5)&=\quad\pmatrix{ 0&0&0&2&3&6   \cr 0&0&9&54&225&882   
\cr 0&6&69&424&2535&14796  \cr}&(B.25)}$$

It is now illustrative to identify the diagrams that are being counted.
We see that, in our formalism, the counting is efficient: each diagram is
counted essentially just with the multiplicity factors of Fig.~1. The tadpole
diagrams have at most a factor two if they differ from their mirror image. See Fig.~9.
\midinsert\cl{\epsffile{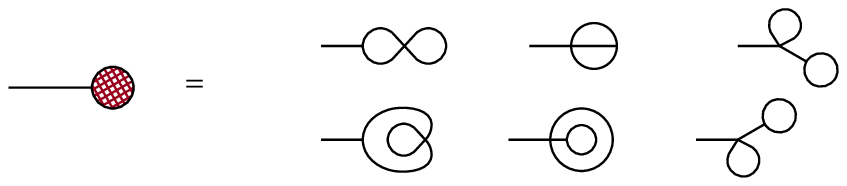}}
\cl{Figure 9. The six diagrams of $F_{0(0)}$ at order $g_0\,\l_0$ (see Eq.~(B.7).}
\endinsert
The two-point Green functions are separated in various classes by our scheme.
In Fig.~10, we see the 48 diagrams of $F_{0(1)}$ at order $g^2\l$,
of which 15 belong to $F_{1(1)}$, and only 5 of those are in $F_{21}$ (cf. Eqs.~(B.8),
(B.13) and (B.15)).
\midinsert\cl{\epsffile{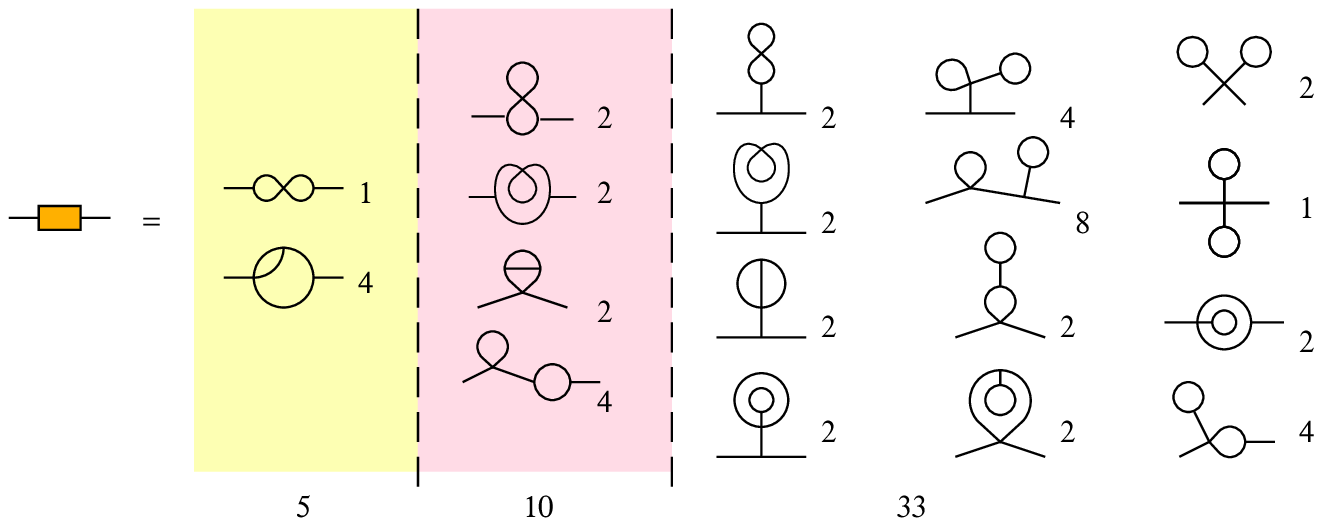}}
\Narrower Figure 10. The 48 diagrams contributing to the 2 point function,
at order $g^2\l$.
15 of these have no tadpole insertion; and 5 have also no seagull insertion.
\Endnarrower\endinsert
Similarly, we illustrate the contributions of different diagrams to the
irreducible 5-point function, to order $g^5\l$ (see Fig.~11). We see
from Eqs.~(B.19), (B.22) and (B.24) that
$\G_{3(4)}$ has 65 entries, $\G_{4(4)}$ has 35, and $\G_{5(4)}$ has
only 5. The diagrams are shown in Fig.~11. Note that the distinction
is whether there are non-trivial irreducible 3-point subgraphs
present, or irreducible 4-point subgraphs.
\midinsert\cl{\epsffile{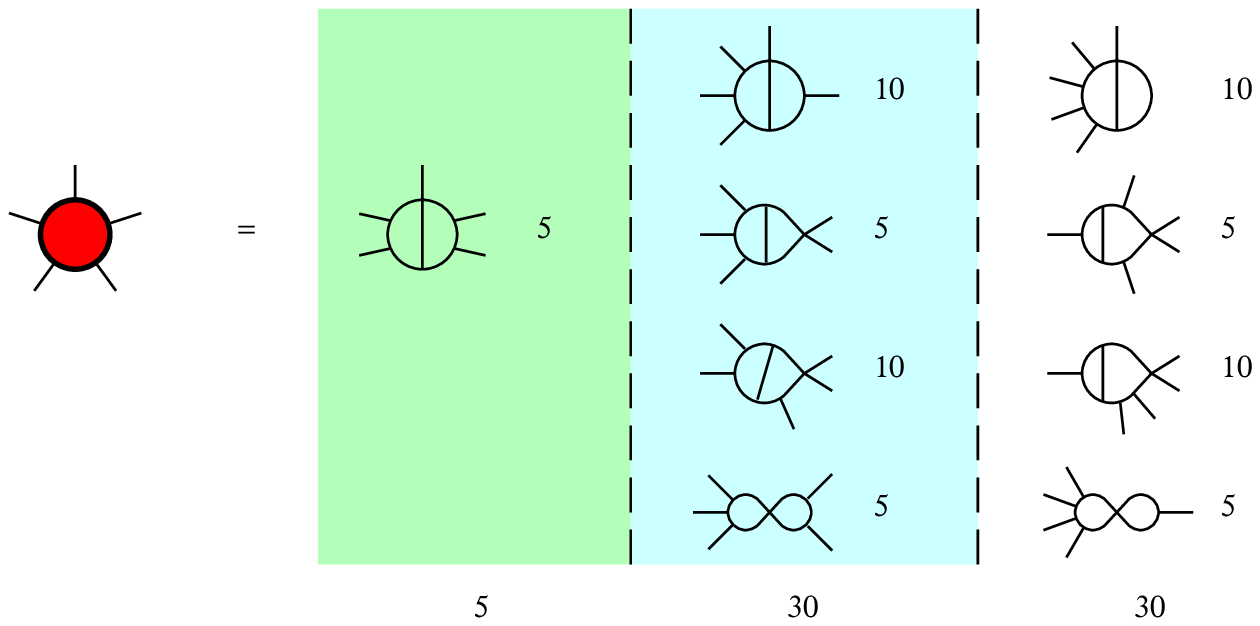}}
\cl{Figure 11. The 64 diagrams of $\G_{3(4)}$, at order $g^5\l$, of which 35 are in
$\G_{4(4)}$ and 5 in $\G_{5(4)}$.}\endinsert

\secbreak
{\ni\bf Appendix C. The critical lines.}\nc
The critical lines (Fig.~7) are found by substituting Eqs.~(9.2) in all our expressions. 
Here we give the outcomes. Starting from Eqs.~(3.16)--(3.21), one just
plugs in the equations (5.7), (5.9), (6.4), (6.5), (7.4), (7.5),
(8.3) and (8.6), and uses the values of $F_{0(i)}$ as they follow from
Eq.~(3.2). We ignore the minus sign in (9.3), which is trivial.
The results are the following algebraic expressions.

\def\tta{6-6\,t-3\,t^2+2\,t^3}
\def\ttb{12-4\,t^3+t^4}
\def\ttc{2+2\,t-t^2}
\def\ttd{144 - 288\,t + 192\,{t^3} - 168\,{t^4} - 48\,{t^5} + 100\,{t^6} - 8\,{t^7} - 17\,{t^8} + 
4\,{t^9}}
\def\tte{ 336 - 672\,t + 384\,{t^2} + 256\,{t^3} - 712\,{t^4} + 48\,{t^5} + 308\,{t^6} - 56\,{t^7} - 
45\,{t^8} + 12\,{t^9}}
\def\ttf{ 24 - 72\,t - 12\,{t^2} + 56\,{t^3} - 10\,{t^4} - 10\,{t^5} + 3\,{t^6}}

$$\eqalignno{g_0^c&=2\,t^2\,(3-t)(\tta)^{1/2}(\ttb)^{-3/2}\,;\onder&(C.1)\cr
\l_0^c&=(\ttc)(\tta) (\ttb)^{-2}\,.&(C.2)}$$	 
$$\eqalignno{g_1^c&={2\,t^2\,(6-t^2)^2(\tta)^{5/2}\over 3^{3/2}\,(\ttd)^{3/2}\onder}\,;\ \ {\ }&(C.3)\cr
\l_1^c&={(\ttc)(\tta)^5\over 9\ (\ttd)^2}	\,.&(C.4)}$$			 
$$ g_2^c={2\,t^2\,(6-t^2)^2(\tta)^{5/2}\over (\tte)^{3/2}\onder}\,;\eqno(C.5)$$
$$\l_2^c={(\ttc)(\tta)^5\over(\tte)^2 }	\,.\eqno(C.6)$$
$$\eqalignno{g_3^c&={2\,t^2\,(6-t^2)^2(\ttf)^{3/2}\over(\tta)^5\onder}	\,;& (C.7)\cr
\l_3^c&={(\ttc)(\ttf)^2\over(\tta)^5}	\,.&(C.8)}$$
$$\eqalignno{g_4^c&={2\,t^2\,(6-t^2)^2(6-3\,t^2+t^3)\over(\ttf)^{3/2}}	\,;\onder&(C.9)\cr
\l_4^c&=\l_3^c	\ .&(C.10)}$$			 
$$\eqalignno{g_5^c&=g_4^c	\ ;\onder&(C.11)\cr
\l_5^c&= \big(3456 - 31104\,t + 25920\,{t^2} + 55296\,{t^3} - 180576\,{t^4} - 44064\,{t^5} \, + &\cr 
&\quad +\, 
  247824\,{t^6} - 27936\,{t^7} - 147672\,{t^8} + 52760\,{t^9}+ 38076\,{t^{10}} - 
  24432\,{t^{11}}\, - &\cr&\quad -\,1762\,{t^{12}}+4506\,{t^{13}} - 921\,{t^{14}} - 202\,{t^{15}} +
114\,{t^{16}} - 18\,{t^{17}} + {t^{18}}
 \big)\,\times\ &\cr
   &\vphantom{\bigg)}\qquad\qquad(\ttf)^{-3}\, .&(C.12)}$$
The curves of Fig.~7 are obtained by plotting $\l_a^c$ against $g^c_a$, 
using $t$ as a parameter.

In all plots, $t$ runs from $1-\sqrt3$ to 0, except for $a=5$, where $t$
runs from $-.890145$ to 0. The new lower bound is the closest zero of
Eq.~(C.12) for $\l^c_5$. The physical reason why this larger domain is needed is easy to
understand: removing the irreducible 4-point diagrams that only contain 3-point vertices
can only happen by invoking a negative counter term in $\l$.
\bye